\UseRawInputEncoding 
\documentclass{resonance}
\usepackage{enumitem}
\usepackage{hyperref}
\usepackage[utf8]{inputenc}
\usepackage[final]{changes}
\begin{document}

%%paper title
%For line breaks, \\ can be used within title 
%The Betelgeuse Enigma Resolved: Betelbuddy exists!}
\title{The Betelgeuse Enigma: The Betelbuddy Hypothesis}
\secondTitle{}
\author{Priya Hasan\footnote{Orcid Id
0000-0002-8156-6940}}

\maketitle
%%\authorIntro is used to place the author's photo and an introduction about the author
%%photo goes into the includegraphics with width=2cm 
%%and a "\\" dividing the text and photo
%%the intro text box is drawn automatically
%%place \authorIntro  just before abstract
\authorIntro{\includegraphics[width=2cm]{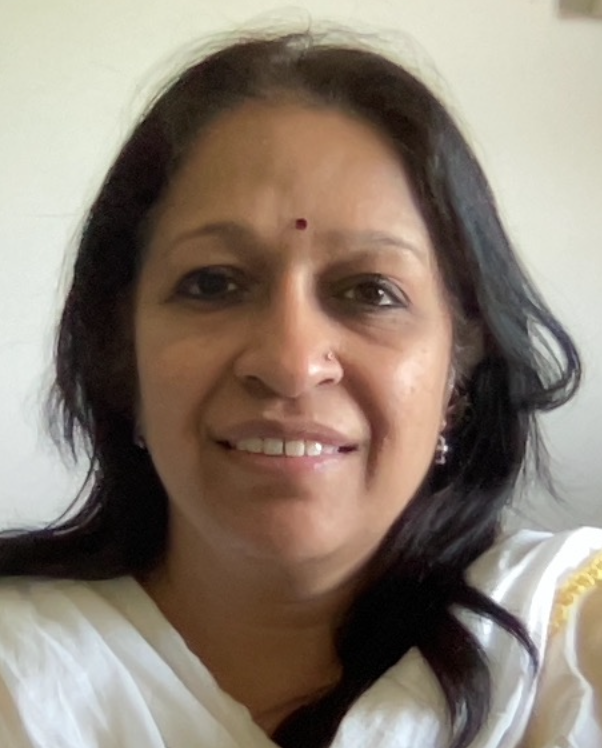}\\
Dr. Priya Hasan is an Assistant Professor of Physics at Maulana Azad National Urdu University, Hyderabad. Her areas of interest are  observational astronomy, star formation, star clusters, and galaxies. Beyond her research, Dr. Hasan is deeply committed to astronomy education and public outreach. She serves as the Regional Astronomy Education Coordinator for the IAU's Office of Astronomy Education (OAE) in India and is a former co-Chair of the IAU's Women in Astronomy Working Group. }
%%abstract
\begin{abstract}
 % Betelgeuse has been in the news for the past six years. In October 2019, it began to dim noticeably, and by mid-February 2020, its brightness dropped by a factor of 3, from magnitude 0.5 to 1.7. This was the time when the excitement that Betelgeuse was ready to go supernova as a 'lull before the storm' was at its peak. In an earlier article by the author, the dimming of Betelgeuse was explained by the episodic mass loss that let out large-sized grain particles that obscured the light from Betelgeuse, making it appear dimmer. This implied that the dimming was not a prelude to the supernova, as it was probably too early for Betelgeuse to go supernova (as estimated)!
 In the past six years, Betelgeuse has been in the news and drawn significant public interest. Starting in October 2019, Betelgeuse underwent a striking dimming event, fading from magnitude 0.5 to 1.7 by mid-February 2020—a threefold decrease in brightness! This gave rise to a number of speculative debates that the star was on the verge of a supernova, a moment of eerie quiet before the cosmic outburst. In a previous article 
 %\cite{hasan20}
 , the author discussed the most accepted explanation to the dimming caused by  episodic mass loss that released large dust grains, which obscured Betelgeuse’s light and made it appear fainter. This interpretation suggested that the dimming was not a precursor to a supernova, as Betelgeuse is likely still far from reaching that stage\footnote{This is debatable, as there are various estimates of when Betelgeuse will go supernova, ranging from anytime now to a 100,000 years.}.

Surprisingly, after that, Betelgeuse gradually brightened up in April 2023 with a peak of  0.1 $V$-band magnitude  and later dimmed to about  0.7 $V$-band magnitude in 2024. 
%A recent study of Betelgeuse, based on stellar models, suggests that Betelgeuse is likely in the carbon-burning phase and istherefore close to a supernova explosion and is a good candidate for the next Galactic Type II supernova in a century or  even a decade.
Betelgeuse  has two periods of variability: a short one of about 400 days, and a longer one, lasting around six years.
 Based on  archival data, two independent studies proposed that the longer period, probably the Long Secondary Period (LSP),  is  due to an undetected low-mass binary companion. Also, radial velocity and astrometric data of Betelgeuse also points to a  companion.  
 Recent observations of Betelgeuse  made with the Atacama Large Millimeter/submillimeter Array (ALMA)  showed that Betelgeuse is rotating at around 5 km$^{-1}$. \keywords{$\alpha$ Orionis, Betelgeuse, Milky Way, supernova, betelbuddy, companion  star, $\beta$ Orionis} It was proposed that Betelgeuse’s high rotation rate is caused by a  merger with a lower mass  companion or gravitational interactions  with a companion which could `spin up' the larger star.  
 HST and Chandra observations were made to find the companion, which were unsuccessful.   On 21 July 2025, the existence of the companion star was confirmed with direct imaging,  for the first time, by the `alopeke' instrument on the 8.1 m Gemini North telescope. The companion is 6 magnitudes fainter than Betelgeuse and orbits close to Betelgeuse itself, within the supergiant star’s extended outer atmosphere. The  stellar companion was detected at an angular separation of 52 mas and a position angle of 115° east of north. These measurements, along with it's brightness,  being roughly 6  magnitudes fainter than Betelgeuse at 466 nm, are in excellent agreement with dynamical predictions. Although this was only a $1.5\sigma$ detection, the agreement in the companion's appearance, separation, position angle, magnitude difference, and estimated mass is reasonable, making the result tentatively acceptable. The next optimum period for observations is November 2027, when we hope to get better observations. The companion star is of the same age as Betelgeuse, 10 million years, but on a totally different stage of evolution, as it differs in mass. While Betelgeuse is in the later stage of evolution as a Red SuperGiant, this is a star not yet born, a pre-main sequence star. However, its fate is sealed: Betelgeuse will cannibalise it in 10,000 years. This article describes the interesting twists and turns  of this study and what lies in the future for Betelgeuse and its companion, Betelbuddy or Siwarha!
\end{abstract}

%%include \monthyear{month year} for month and year of publication in the footer
\monthyear{November 2024}
%%use \artNature for the running head information
\artNature{GENERAL  ARTICLE}

\begin{center}
\it{The cause is hidden. The effect is visible to all.
― Ovid.}

 \end{center}

\section{Introduction}
\begin{figure}[!t]
\caption{A schematic of the Orion constellation, with a small pink arrow indicating the location of Betelgeuse in Orion’s left shoulder. Image credit: Wikimedia Commons}
\label{betel} 
\vskip -12pt
\centering
\includegraphics[width=9.0cm, height=10.0cm]{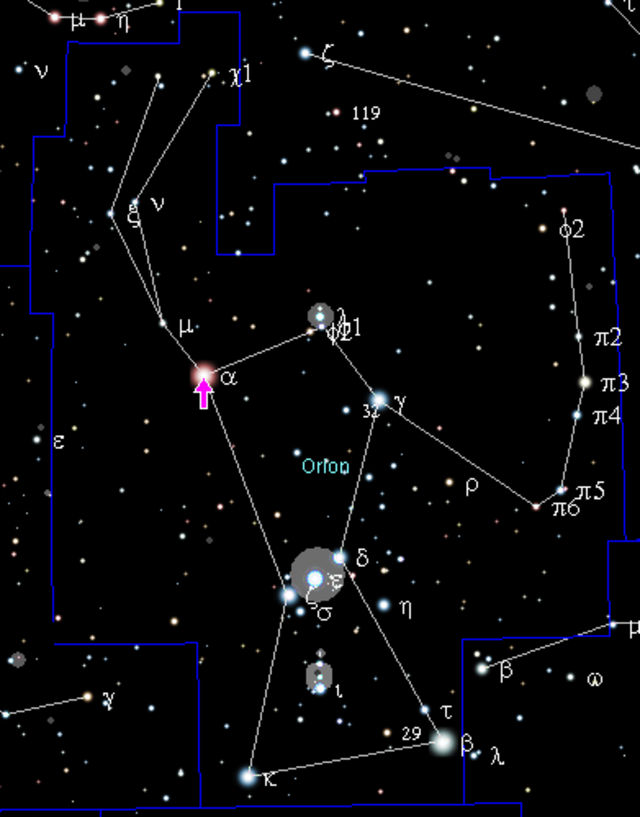}
\end{figure}

 Betelgeuse ($\alpha\ Ori$ A, HD 39801) is no ordinary star. It is one of the best-known and nearest red supergiant (M2Iab) star to Earth and reasonably easy-to-locate  in the constellation Orion (Fig.~\ref{betel}).  It  has long been known to have modulations in both visible light and radial velocity, coupled with semiregular fluctuations  with four approximate periods of \deleted{approximately} 2200, 420, 230, and 185 days.

\begin{figure}[!t]
\caption{The four-panel graphic explains the Great Dimming with a dust cloud. A hot blob of plasma was likely ejected from a large convection cell on the star. As the gas rapidly expanded away from the star, it cooled and turned into a giant dust cloud. This dust cloud blocked the light from a quarter of the star's surface when seen from Earth. Credit: NASA/ESA/Elizabeth Wheatley (STScI).}
\label{beteld} 
\vskip -12pt
\centering
\includegraphics[width=9.0cm, height=4.0cm]{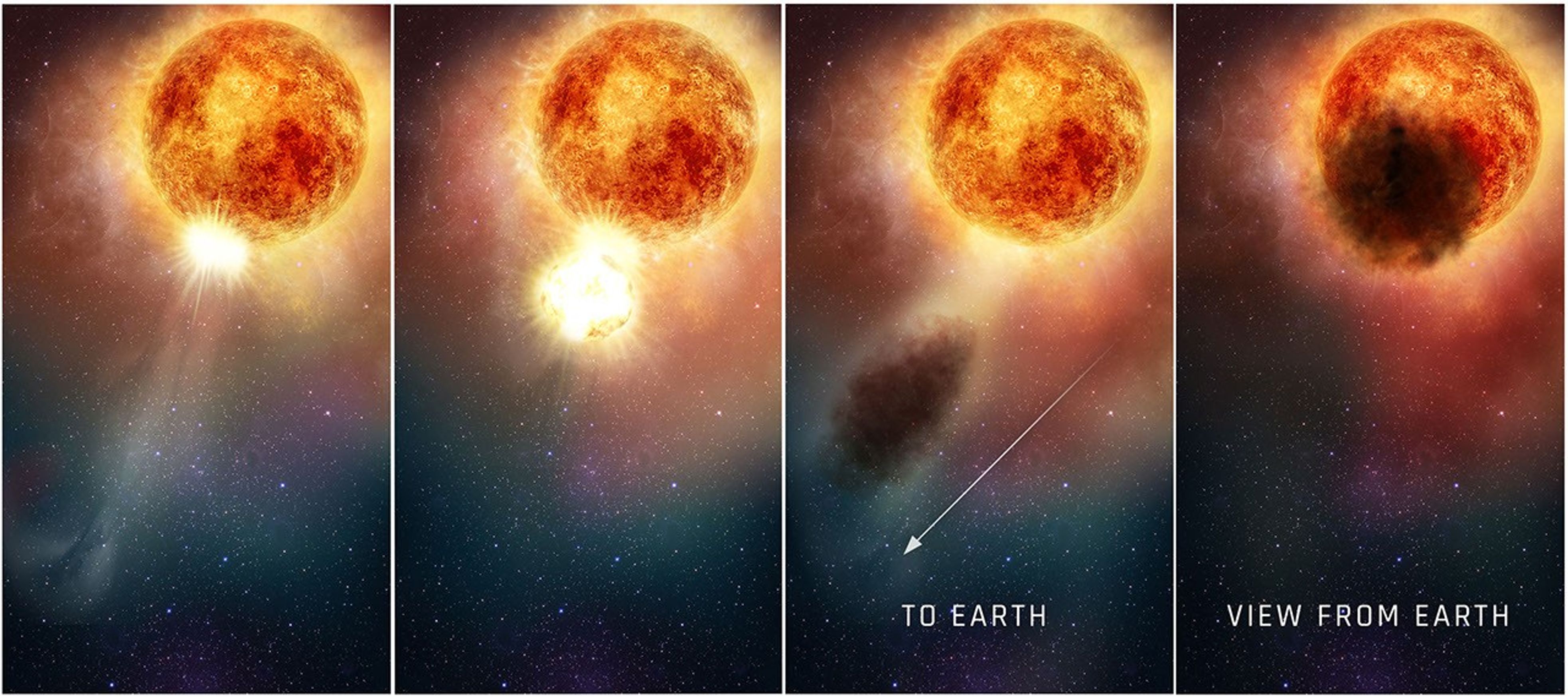}
\end{figure}

\rightHighlight{\textbf{The Great Dimming}\\ 
Starting in October 2019, Betelgeuse underwent a striking dimming event, fading from magnitude 0.5 to 1.7 by mid-February 2020—a threefold decrease in brightness! People were excited that it may go supernova, but finally the most acceptable explanation was a  Surface Mass Ejection (SME)that caused a large dust cloud in our line of sight.}Interest in Betelgeuse has increased over the past six years, initiated by an anomalous dimming in its brightness observed between the end of 2019 and the beginning of 2020 \cite{gui19}-\cite{dup20}. It is now widely accepted that a dust cloud  or cooling of its surface due to convection changes was the cause \cite{2021Natur.594..365M, hasan20} (Fig. \ref{beteld}). The `Great Dimming' nonetheless inspired a fresh interest in the star and  led to a novel understanding of Betelgeuse and it's  parameters\footnote{The reader is advised to read an earlier paper by the author \cite{hasan20} on Betelgeuse to follow the complete development of our present understanding of the `Great Dimming' of the star.}.

Astronomers have concluded that the red supergiant Betelgeuse `blew its top' in a cataclysmic event, never before seen in a normal star. By analyzing data from NASA's Hubble Space Telescope and other observatories, they determined the star ejected a substantial part of its visible surface in a gigantic Surface Mass Ejection (SME) (Fig. \ref{beteld}).

This event dwarfs our Sun's coronal mass ejections (CMEs); the Betelgeuse SME blasted away 400 billion times more mass than a typical solar CME!

The story with Betelgeuse has further unfolded with the first direst observation of a companion star to explain Betelguese's strange behaviour. 
\section{Betelguese: A close look}
\leftHighlight{\textbf{Fast Facts on Betelguese:}\\ Distance:  214 pc \\ Age: 7--11~Myr\\Progenitor Mass:  $17-19 M_{\odot}$ \\ Radius $700 R_{\odot}$ \\ Rotation:  17 years\\ Temperature: 3600 K
Luminosity 100,000 L$_{\odot}$\\
}
Let's begin by compiling the key facts about Betelgeuse. It is a red super giant (RSG) \deleted{at}  with its parameters listed on the left. 

\begin{figure}[!t]
\caption{Betelgeuse  imaged by   Atacama Large Millimeter/submillimeter Array (ALMA), in the millimeter continuum. The star  is $\approx $ 1400 times larger than our Sun. The orbital paths of the Solar System planets are shown for comparison.
Crédit: ALMA (ESO/NAOJ/NRAO)/E. O’Gorman/P. Kervella}
\label{alma} 
\vskip -12pt
\centering
\includegraphics[width=7.0cm, height=7.0cm]{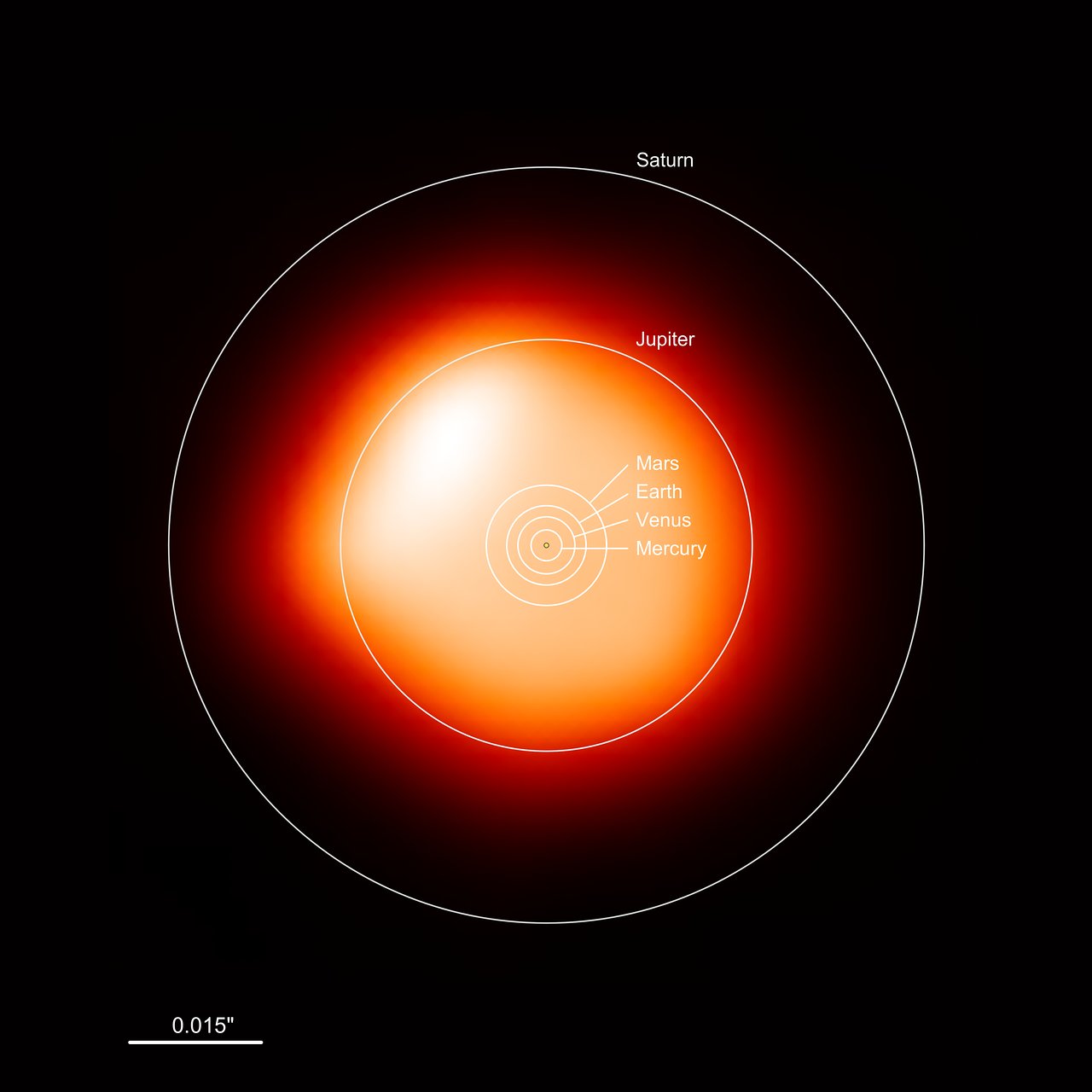}
\end{figure}

Since Betelgeuse is big and near, we can resolve its surface with ground or space-based telescopes. Betelgeuse was the first star, other than the Sun, to have a direct measurement of its angular diameter (47 mas
at 575 nm \cite{Michel21}\added{)}. We have imaged the star in various wavelength and also discovered surface features that have helped understand the processes taking place (Fig. \ref{alma}). 
\cite{dolan16} estimated that Betelgeuse will explode as a Type II supernova in less than 100,000 years. This cataclysmic event will release a staggering $2 \times 10^{53}$ erg in neutrinos and $2 \times 10^{51}$ erg as kinetic energy, leaving behind a neutron star with a mass of $1.5 M_{\odot}$. By this time, the star will have evolved \added{as} a silicon-iron core. When it explodes, it will be the closest observed supernova in history, occurring at a distance 19 times nearer than Kepler's Supernova\added{\footnote{Kepler's Supernova appeared in 1604  and  is the most recent supernova in the Milky Way galaxy to have been  observed by the naked eye}}. \\

% Due to expansion and hence conservation of angular momentum, giants tend to rotate slowly.  Betelgeuse  spins  once every 8.4 years, whereas the Sun spins once a month. Luminosity is directly proportional to the radius $R$ and temperature $T$ of a star and is given by:
%  $$L=\sigma 4 \pi R^2T^4$$

% Due to its large size, albiet it's lower temperature, its luminosity is $125,000 \times  L_{\odot}$. Betelguese has used up the hydrogen fuel in it's core and is presently fusing helium into carbon, which is very efficient and hence generates lot of heat. In the process, it blows a very strong wind of material away from it $\approx  10^{-6} M_{\odot}  yr^{-1}$. In comparison the Sun loses $< 10^{-12} M_{\odot} yr^{-1}$.

\section{Variability of Betelguese}
\rightHighlight{A variable star is a star whose brightness varies. This  may be caused by a change in emitted light or by something partly blocking the light. Variable stars are classified as either: Intrinsic variables, whose luminosity changes due to physical processes on the star or extrinsic variables, whose apparent changes in brightness are due to changes in the amount of their light that can reach Earth; for example, in eclipsing binaries. \\
Most stars do vary in their luminosity. The energy output of our Sun, varies by about 0.1\% over it's 11-year solar cycle} 
Classified as a pulsating red supergiant (RSG), Betelgeuse is unstable and undergoes radial pulsations. These expansions and contractions result from a significant imbalance between outward radiation pressure from nuclear fusion and inward gravitational force. The amplitude of these pulsations is immense; the star's radius at minimum size reaches the orbital distance of Mars, while at maximum expansion, it swells to a radius approaching Jupiter's orbit (Fig. \ref{alma}).

\begin{figure}[!th]
\caption{AAVSO photometric data in the V and B bands traces Betelgeuse's brightness over the last eight years, capturing the profound Great Dimming event of late 2019 to early 2020. The timeline also marks the dates of two speckle observations. (Image Credit: AAVSO)}
\label{var} 
\vskip -12pt
\centering
\includegraphics[width=12cm, height=7cm]{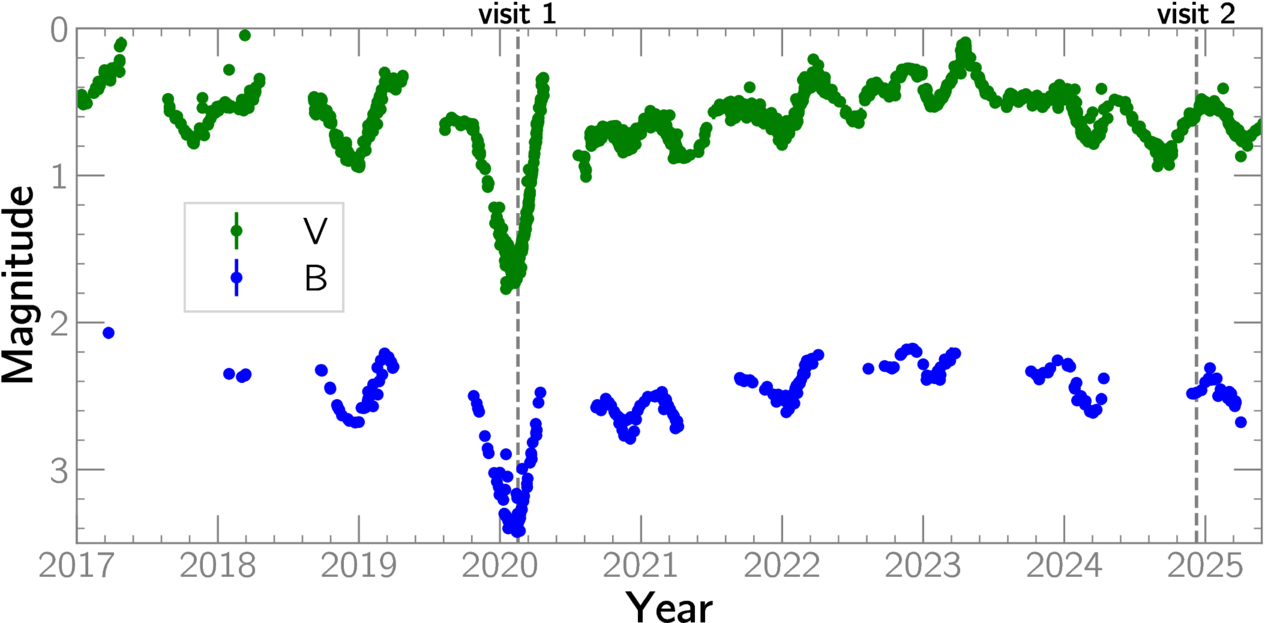}
\end{figure}

Betelgeuse exhibits semi-regular variability driven by multiple periodic oscillations. \cite{2023MNRAS.526.2765S} identified four distinct modes with periods of 2190 $\pm$ 270, 417 $\pm$ 24, 230 $\pm$ 29, and 185 $\pm$ 4 days (Fig. \ref{var}). Additional brightness fluctuations arise from surface inhomogeneities, including huge starspots (analogous to sunspots) and bright regions of convecting plasma. Currently in a bright phase, Betelgeuse will ultimately exhaust its nuclear fuel, undergo gravitational collapse, and explode as a Type II supernova.
Betelgeuse has 2 pulses. One pulse is about 6 years, while the second is little more than a year. One of these pulses is the so-called “fundamental mode” – a property of the star itself, related to pulsations. If the fundamental mode is the longer-scale pulse, then Betelgeuse could go supernova sooner than expected. Previous studies suggest that the fundamental mode of Betelgeuse is the shorter heartbeat. In that case, the Long Secondary Period (LSP)  must be due to external factors such as the presence of a companion, Betelbuddy.

 % Astronomers Edward Guinan and Richard Wasatonic (Villanova University) and amateur astronomer Thomas Calderwood, had been monitoring the star for more than 25 years. They reported a fall to magnitude 1.29 on December 20th  2019 using precise V-band photometry, this being the lowest in records. In Astronomical Telegram  No. 13365 Guinan concluded that  `The current faintness of Betelgeuse appears to arise from the coincidence of the star being near the minimum light of the 5.9~yr light-cycle as well as near the deeper than usual minimum of the 425~day period.' In effect, the star's overlapping cycles have created a sort of superminimum. 
\leftHighlight{The time periods for the various Fusion Processes in a star are:\\	
Hydrogen to Helium:	10 million years\\
Helium to Carbon: 	500,000 years\\
Carbon to Neon:	600 years\\
Neon to Oxygen:	1 year\\
Oxygen to Silicon:	6 months\\
Silicon to Iron:	1 day}

\cite{2023MNRAS.526.2765S} proposed a controversial model centered on the interpretation of Betelgeuse's LSP as the fundamental mode (FM) of radial pulsation, suggesting Betelgeuse is further along in its stellar evolution than previously thought, potentially indicating an imminent supernova.  If the LSP periodicity reflects an intrinsic variation associated with Betelgeuse’s fundamental (acoustic) pressure mode (FM), it would suggest a stellar radius exceeding 1000 R$_{\odot}$ and indicate that the star is in its core carbon-burning phase—a brief evolutionary stage lasting only a few hundred to a few thousand years.  The analysis suggests Betelgeuse is nearing the end of its core carbon-burning phase. With only about 600 years of carbon fuel remaining, the subsequent core collapse and supernova explosion are expected to occur rapidly, meaning the star is on the verge of exploding in astronomical terms.

However, \cite{2023RNAAS...7..119M} pointed out that the angular diameter measurements of the star are in conflict with the stellar radius found by them ($764^{+116}_{62}R_{\odot}$). They proposed  that the Great Dimming was caused by constructive mode interference using long-term brightness measurements and that Betelgeuse's $\approx$400 day period is the result of pulsation in the fundamental mode, driven by the $\kappa$-mechanism.

The characterization of its variability was also studied by \cite{2024A&A...685A.124J}. They used extensive high-resolution spectroscopic data  with 
\rightHighlight{\textbf{Evidence of binarity}\\  It seems that the Betelgeuse's LSP is not the fundamental mode and so binarity seems the best explanation \\ The large rotation rate of Betelgeuse  5 km s$^{-1}$ suggestions that Betelgeuse is a binary merger.\\ Radial velocity, visual magnitude, and astrometric data,  of more than a century supports the presence of a low-mass companion  \\ HST and Chandra observatons provided constraints on the companion  \\Estimates made of the position, the appearance of the companion; the angular separation from Betelgeuse; the position angle with respect to Betel-
geuse; the magnitude difference; and the estimated mass of the
companion }the tomographic method and concluded that powerful shocks were the triggering mechanism for episodic mass-loss events, the most accepted explanation of the the Great Dimming. \cite{joy24}  demonstrated that models of single-star evolution are insufficient to reproduce Betelgeuse's present-day rotational velocity, implying a merger event or a companion as a  source of angular momentum is required.  The analysis also definitively located Betelgeuse on the red supergiant branch in the early core helium-burning phase and constrained its present-day mass to 16.5–19 M$_{\odot}$, revising previous literature values downward.
 
\section{Evidence of binarity}

For resolved objects like the Sun, planets in our Solar System are easy to measure the rotation speed by observing the movement of a feature on the surface. The Sun rotates at a speed of 2 kms$^{-1}$, but is a thousand times smaller  and ten times less massive than than Betelgeuse. If the Sun were to expand to the size of Betelgeuse, it would reach the orbit of Jupiter and its surface velocity would be 2 ms$^{-1}$. If the Sun had the mass of Betelgeuse, its surface velocity would be 0.2 ms$^{-1}$. Cool evolved stars generally don't spin very fast, at least not on the outside. The stars' orbits get bigger by one or two orders of magnitude as they change over time. Because rotational momentum is conserved, this causes the outer layers to slow down. 

\begin{figure}[!t]
\caption{Graphical depiction of Betelgeuse and the Betelbuddy. Credit: Lucy Reading-Ikkanda/Simons Foundation}
\label{nr} 
\vskip -12pt
\centering
\includegraphics[width=6cm, height=6 cm]{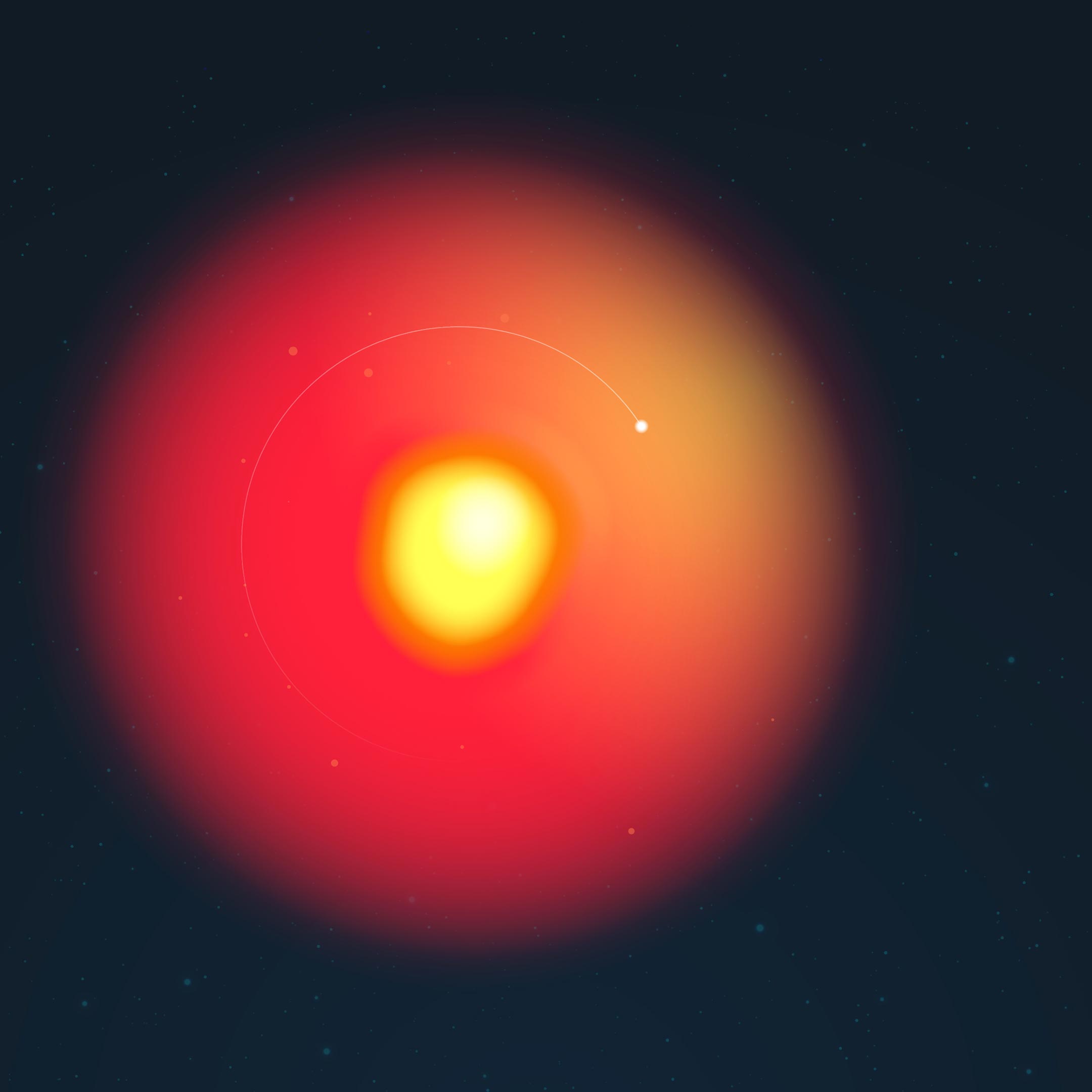}
\end{figure}

And now the surprise: A new study using the Atacama Large Millimeter/Submillimeter Array (ALMA) \cite{Ma24} found a  projected rotation rate of Betelgeuse of about 5 km s$^{-1}$. This value is two orders of magnitude larger than single-star evolution models predict, strongly suggesting that Betelgeuse is the product of a binary merger or has gained angular momentum from a companion.
  However, the same paper, \cite{Ma24} proposed that large-scale convective motions, especially if they are only partly resolved, can look like rotation, instead. They showed, through simulations, that in 90\% of the time, rotation could be confused with turbulent convective flows. If that's true,  a binary companion may not exist. Higher resolution images of Betelgeuse would help resolve this dichotomy. 
\leftHighlight{The fundamental mode of vibration is the simplest, lowest-frequency pattern in which a system naturally oscillates or vibrates, often referred to as the first harmonic or the lowest resonant frequency. This basic mode has a specific shape and frequency for a given system, and other, higher-frequency vibration modes (overtones or harmonics) can be understood as multiples of this fundamental vibration.} 
\cite{mor24} analyzed radial velocity, visual magnitude, and astrometric data,  of more than a century, for Betelgeuse,  to revisit the long-standing hypothesis that it may be a spectroscopic binary. The observations showed that Betelgeuse exhibits stochastic variability over years and decades due to its turbulent convective envelope, a periodic secondary variation with a 5.78-year LSP, and quasiperiodic pulsations on timescales of several hundred days. The LSP is consistently detected in both astrometric and radial velocity measurements, supporting the presence of a low-mass companion, likely less than one solar mass, orbiting with a 2110-day period at a distance slightly more than twice Betelgeuse’s radius. \cite{2025arXiv250518375G} asserted that the Long Secondary Period (LSP) in Betelgeuse's variability is caused by a low-mass binary companion, provisionally named $\alpha\ Ori$ B. This companion would be about 20 times less massive and a million times fainter than Betelgeuse, with a similar temperature, rendering it effectively invisible (Fig \ref{nr}). The authors based this on a far-ultraviolet campaign with HST/STIS, searching for the companion's spectral signatures. Their astrometric evidence suggests an edge-on binary orbit aligned with Betelgeuse’s spin axis. They propose that tidal spin-orbit coupling transfers angular momentum from the orbit to the star, explaining both this alignment and Betelgeuse's rapid rotation. Ultimately, orbital decay is expected to lead to the companion's engulfment within the next 10,000 years.\footnote{Papers that predicted Betelgeuse's companion believed that no one would likely ever be able to image it \cite{howell2025}}. 
\cite{{grady25}} put X-Ray Constraints on the Nature of $\alpha \  Ori$ B to identify any X-ray emission from the companion.
Both the above studies, based on HST and Chandra data,  $\alpha\   Ori$ B is most consistent with a Young Stellar Object (YSO) with a mass 0.5 - 1.1M$\odot$. 
%By comparing our data with canonical YSO spectra from the ULLYSES database, we rule out companion masses ≳ 1.5 M⊙ and far-UV (≈1200–1700 Å) continuum or line emission stronger than ≈ 10⁻¹⁴ erg s⁻¹ cm⁻² Å⁻¹.
Future campaigns coordinated with the LSP phase would be essential to further constrain the spectroscopic properties of $\alpha\   Ori$ B.
\begin{center}
\begin{figure}[!th]

\label{bb2} 
\vskip -12pt
\centering
\includegraphics[width=14cm, height=8 cm]{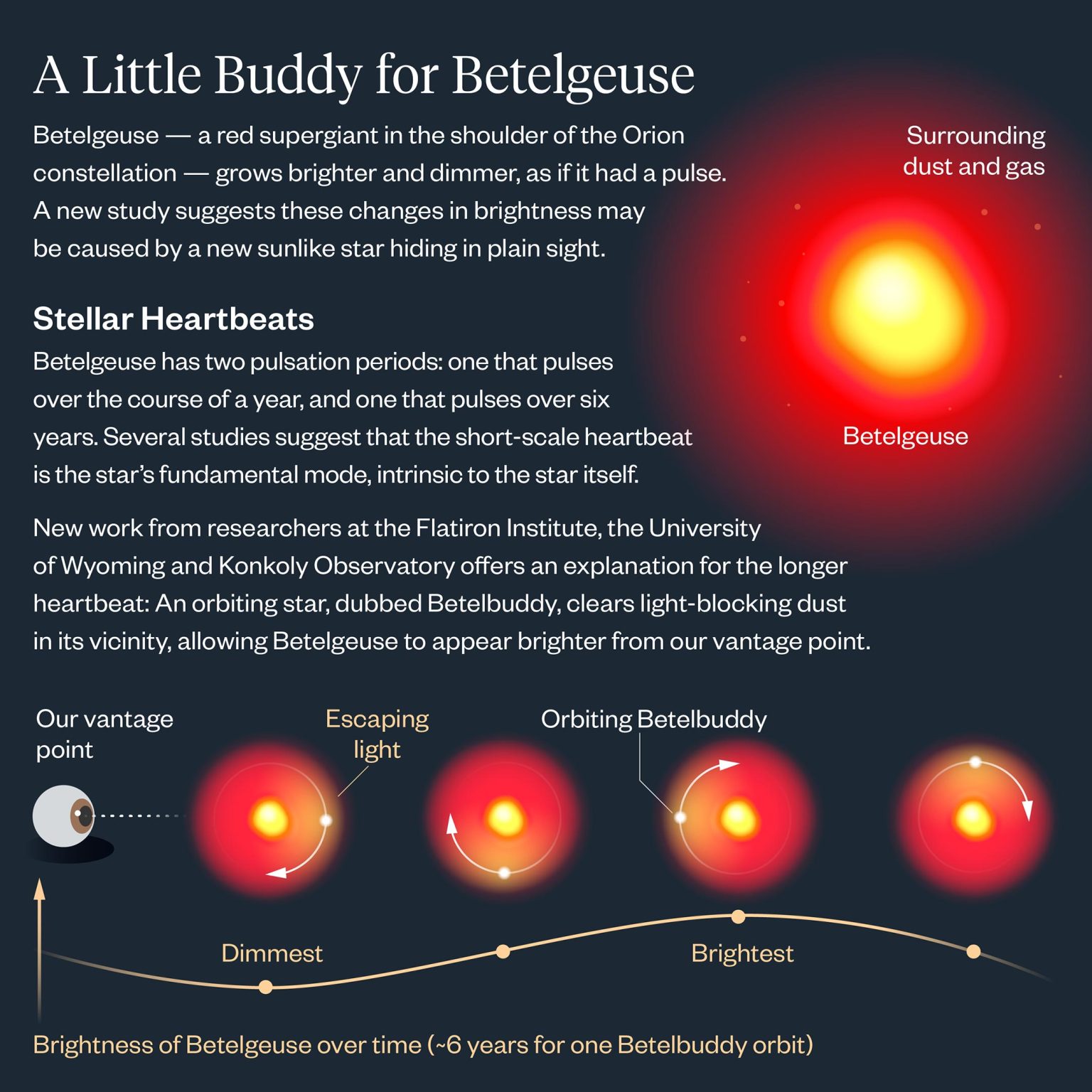}
\caption{Infographic explaining Betelbuddy and its affect on the brightness of Betelgeuse. Credit: Lucy Reading-Ikkanda/Simons Foundation}
\end{figure}
\end{center}
Figure \ref{bb2} describes the results of previous studies that has very accurately calculated the position, mass and alignment of Betelbuddy. The next step was to verify its existence, though it was deemed to be very difficult.   

\section{Direct Imaging}

Observations of the companion of  Betelgeuse are difficult because it orbits Betelgeuse in an incredibly tight orbit and the brightness of Betelgeuse would make it impossible to detect this faint source, so near. 

\begin{figure}[!t]
\caption{Observations of Betelgeuse and its companion star as seen for the first time by the `alopeke' instrument on the Gemini North telescope in December 2024. (Image credit: International Gemini Observatory/NOIRLab/NSF/AURAImage Processing: M. Zamani (NSF NOIRLab))}
\label{bb3} 
\vskip -12pt
\centering
\includegraphics[width=9cm, height=5 cm]{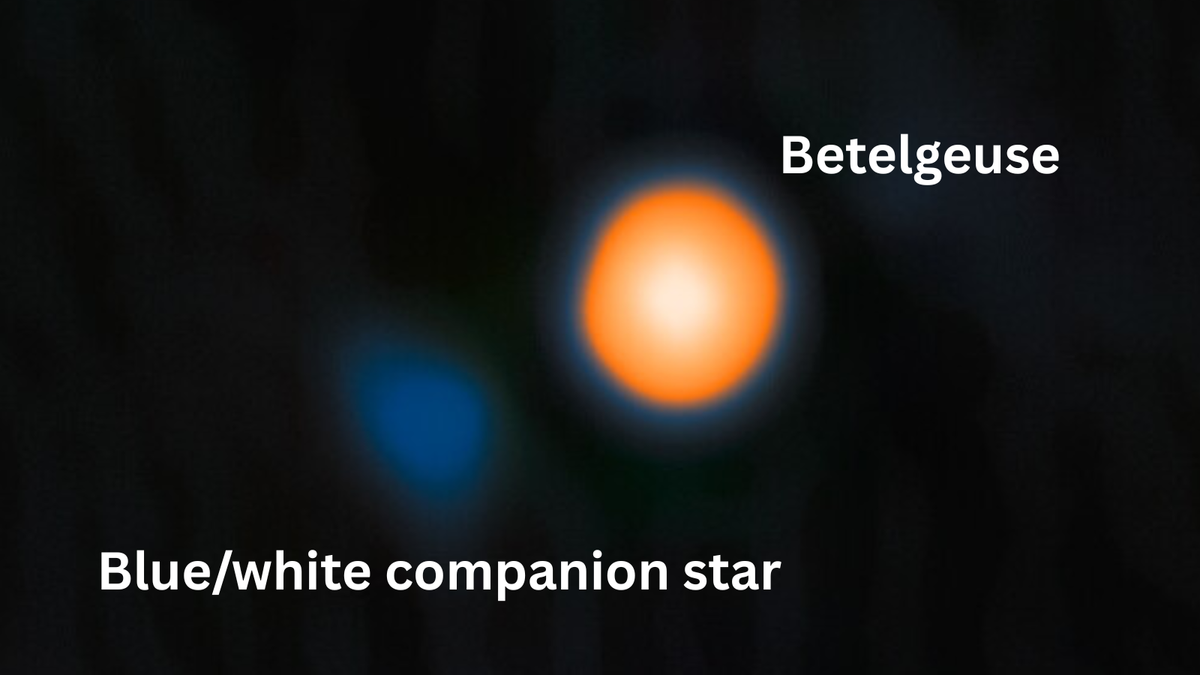}
\end{figure}

\leftHighlight{Speckle imaging  is a technique used to overcome the blurring effects of the Earth's atmosphere on ground-based optical telescopes. It involves taking numerous short exposures to compensate for atmospheric distortions and then analyzing them to reconstruct a high-resolution image. %The atmosphere limits the resolution of ground-based telescopes, and speckle imaging helps to mitigate this limitation, allowing for more detailed observations of celestial objects. 
}
A team of astrophysicists led by NASA Ames Research Center scientist Steve Howell investigated Betelgeuse using the Gemini North telescope and its 'Alopeke speckle imager \cite{howell2025}. On 9 December 2024—a date predicted by theoretical calculations to show maximum stellar elongation—the team made the first-ever detection of Betelgeuse's faint companion (Fig. \ref{bb3}). Although this was only a $1.5\sigma$ detection, the observed properties (appearance, angular separation, position angle, magnitude difference, and mass) align reasonably with predictions, making the result tentatively acceptable.

The companion, named Siwarha (Arabic for "her bracelet"), has a mass of approximately 1.5M$_{\odot}$ and is a hot, blue-white pre-main sequence star orbiting within Betelgeuse's extended atmosphere at just four times the Earth-Sun distance. 
\leftHighlight{The highest angular resolution achievable with the Gemini Observatory is $\approx$ 0.04-0.06 arcseconds using the Gemini Multi-Conjugate Adaptive Optics System (GeMS) and its companion imager GSAOI. The Gemini speckle cameras, 'Alopeke' and `Zorro',  reach resolutions of $\approx$ 0.018 arcseconds in the optical bandpass, though this is for single targets.}This proximity makes it the closest companion ever detected around a red supergiant. The authors suggest that while both stars likely formed at the same time, Betelgeuse evolved faster due to its greater mass, whereas the lower-mass companion has not yet ignited hydrogen fusion.

Despite its youth, the companion’s future is precarious. The team predicts Betelgeuse's intense gravity will tidally disrupt or cannibalize the smaller star within the next 10,000 years. Astronomers anticipate another observational opportunity in November 2027, when the companion reaches maximum separation.

Beyond revealing Betelgeuse’s binary nature, this discovery provides new insights into the long-period variability of red supergiants and their complex evolutionary pathways.

\footnote{Betelgeuse means the hand of al-Jawzā’, the authors proposed the name Siwarha for the probable companion star, which means her bracelet in Arabic} This represents the first time a companion star has been detected so close to a red supergiant.

% Eventually Betelguese will go supernova !!! At that distance,  it'll be as bright as the full Moon! Luckily, too far away to hurt us. And that too after 100,000 yrs! Octillions ($10^{27}$) of tons of matter into space in all directions at a decent fraction of the speed of light.  The shock wave will take $6\times 10^6$ yrs to reach us, at 13~kms$^{-1}$.  The shock  will stop  outside the Earth's orbit.
% We're safe.
 
% But poor Orion..... with Betelgeuse gone,  he'll be missing his right shoulder that we are all so familiar with!

\section{Conclusion}
Over the past six years, Betelgeuse has captured the spotlight and fascinated both astronomers and the public. In late 2019, the star underwent a dramatic dimming event, fading from magnitude 0.5 to 1.7 by mid-February 2020—a threefold drop in brightness. This unexpected change sparked widespread speculation that Betelgeuse might be on the brink of going supernova, a moment of hushed anticipation before a cosmic explosion.
\rightHighlight{3$\sigma$ detection refers to using the three-sigma rule to identify anomalies or outliers in a dataset that is assumed to follow a normal distribution. Data points falling outside three standard deviations ($\sigma$) from the mean (M) are flagged as potential outliers because this range typically contains 99.7\% of the data in a normal distribution. This method is widely used in quality control, statistical analysis, and anomaly detection to set thresholds and signal when a process is behaving unexpectedly.}
This dimming was explained by episodic mass loss that produced large dust grains. These dust clouds obscured the star’s light, creating the illusion of fading rather than signaling an imminent supernova. In reality, Betelgeuse is probably still far from that final stage of stellar evolution.

Interestingly, Betelgeuse did not simply return to normal. Instead, it brightened to an impressive 0.1 V-band magnitude in April 2023, only to dim again to about 0.7 V-band magnitude in 2024. This variability aligns with what is already known: Betelgeuse has a primary brightness cycle lasting around 400 days, along with a secondary, longer cycle of roughly six years.
Archival observations have fueled further intrigue. Two independent studies proposed that this Long Secondary Period (LSP) might be driven by a hidden, low-mass stellar companion. Radial velocity and astrometric data also hinted at the presence of such a companion. Finally, on 21 July 2025, direct imaging by the ‘Alopeke instrument on the Gemini North telescope confirmed the existence of this long-suspected companion.

The newly detected star is about six magnitudes fainter than Betelgeuse and lies extremely close, orbiting within the extended atmosphere of the red supergiant. Although the detection is not valid in significance ($1.5\sigma$), \leftHighlight{Betelgeuse's variability, including brightening in 2023 and dimming again in 2024, reflects both its primary ~400-day cycle and a secondary ~6-year cycle.

Archival studies suggested that the longer cycle might be caused by a hidden stellar companion. This was confirmed on 21 July 2025, when the ‘Alopeke instrument on Gemini North directly imaged a faint, low-mass companion within Betelgeuse’s extended atmosphere. Though the detection (at $1.5\sigma$) requires further validation during the next elongation in 2027, its properties match earlier predictions.

The companion, about 10 million years old, is still a pre-main-sequence star, unlike Betelgeuse, which is in its Red Supergiant phase. Ultimately, Betelgeuse will engulf the companion in ~10,000 years. This discovery has been nicknamed Betelbuddy or Siwarha.}the measurements agree well with earlier predictions—its appearance, angular separation, position angle, brightness difference, and estimated mass. The two stars will once again be at their maximum elongation in November 2027 and that's when new observations will be required to validate this result. 
The companion is coeval with Betelgeuse, around 10 million years old, but is at a completely different evolutionary stage. While Betelgeuse is a Red Supergiant nearing the end of its life, the companion is still a pre-main sequence star—a star not yet fully formed. Tragically, its destiny is sealed: within the next 10,000 years, Betelgeuse will engulf and consume it.

This discovery adds yet another fascinating chapter to the story of Betelgeuse. As astronomers continue to watch this stellar giant and its newfound partner—nicknamed Betelbuddy or Siwarha—the system promises more surprises in the years to come.

`This detection was at the very extremes of what can be accomplished with Gemini in terms of high-angular resolution imaging, and it worked,' Howell said. `This now opens the door for other observational pursuits of a similar nature.'
This study also opens up the possibility of RSG stars like Betelgeuse: Antares and Arcturus which both have extra-long variability periods to possibly have a hidden companion as the reason for their variability and further enhance our understanding of these stars.
And finally, this is yet another interesting tale of how research in astronomy is full of twists and turns that make it even more interesting and fun!!!

\section*{Acknowledgements}
%The author would like to thank the referee for her/his valuable comments that helped improve the content of the article. 

%%References section

\end{document}